\documentclass[final,5p,times,twocolumn]{elsarticle}

\usepackage{lineno,hyperref}
\modulolinenumbers[5]

\usepackage{amsmath}
\usepackage{booktabs}

\journal{Nuclear Instruments \& Methods in Physics Research, Section A}

\begin{document}

\begin{frontmatter}

\title{Compensated LGAD optimisation through van der Pauw test structures}

\author[1,2]{A.~Fondacci\corref{cor1}}
    \ead{alessandro.fondacci@pg.infn.it}
    \cortext[cor1]{Corresponding author}
\author[2]{T.~Croci}
\author[3,2]{D.~Passeri}
\author[4,5]{R.~Arcidiacono}
\author[5]{N.~Cartiglia}
\author[5]{M.~Ferrero}
\author[6]{M.~Centis Vignali}
\author[6]{M.~Boscardin}
\author[6]{G.~Paternoster}
\author[5]{R.~S.~White}
\author[5]{A.~R.~Altamura}
\author[7,5]{V.~Sola}
\author[2]{A.~Morozzi}
\author[8,2]{F.~Moscatelli}

\affiliation[1]{organization={Dipartimento di Fisica, Università degli Studi di Perugia},
                addressline={Via Alessandro Pascoli},
                postcode={06123},
                city={Perugia},
                country={Italy}}
\affiliation[2]{organization={Istituto Nazionale di Fisica Nucleare (INFN) - Sezione di Perugia},
                addressline={Via Alessandro Pascoli},
                postcode={06123},
                city={Perugia},
                country={Italy}}
\affiliation[3]{organization={Dipartimento di Ingegneria, Università degli Studi di Perugia},
                addressline={Via Goffredo Duranti, 93},
                postcode={06125},
                city={Perugia},
                country={Italy}}
\affiliation[4]{organization={Dipartimento di Scienze del Farmaco, Università del Piemonte Orientale},
                addressline={Largo Donegani, 2},
                postcode={28100},
                city={Novara},
                country={Italy}}
\affiliation[5]{organization={Istituto Nazionale di Fisica Nucleare (INFN) - Sezione di Torino},
                addressline={Via Pietro Giuria, 1},
                postcode={10125},
                city={Torino},
                country={Italy}}
\affiliation[6]{organization={Fondazione Bruno Kessler (FBK)},
                addressline={Via Sommarive, 18},
                postcode={38123},
                city={Trento},
                country={Italy}}
\affiliation[7]{organization={Dipartimento di Fisica, Università degli Studi di Torino},
                addressline={Via Pietro Giuria, 1},
                postcode={10125},
                city={Torino},
                country={Italy}}
\affiliation[8]{organization={Istituto Officina dei Materiali (IOM) CNR - Sede di Perugia},
                addressline={Via Alessandro Pascoli},
                postcode={06123},
                city={Perugia},
                country={Italy}}

\begin{abstract}
    A new gain implant design has recently been introduced to enhance the radiation resistance of low-gain avalanche diodes (LGADs) to the extreme fluences anticipated in future hadron colliders like FCC-hh. This design utilises an engineered compensation of two opposing types of doping implants, requiring a thorough analysis of their evolution due to irradiation. To this end, the experimental measurements of their initial test structures have been compared with Technology CAD simulations both before and after irradiation.

    From the measurement-simulation comparison regarding C-V characteristics, the donor removal at high initial donor concentrations ($>10^{16}$ at/cm$^3$) used in Compensated LGADs has been studied, along with how donor co-implantation influences the beneficial effect of carbon to slow acceptor removal. Furthermore, an innovative application of van der Pauw test structures, typically employed by foundries to monitor process quality, has been implemented. The doping removal of the single implants used in Compensated LGADs has been estimated by examining the variation in sheet resistance with irradiation through these structures.
\end{abstract}


\end{frontmatter}


\section{Introduction}
\label{sec:Introduction}

    Future hadron colliders, such as FCC-hh, will have increasingly crowded environments, and time will be an essential parameter to be added to event analysis~\cite{Detector:2784893}. Timing resolutions of tens of picoseconds will enable the separation of the tracks in time (4D-tracking), leading to fewer tracks per event and, thus, a more straightforward analysis.

\begin{figure}[h]
        \centering
        \includegraphics[width=0.9\linewidth]{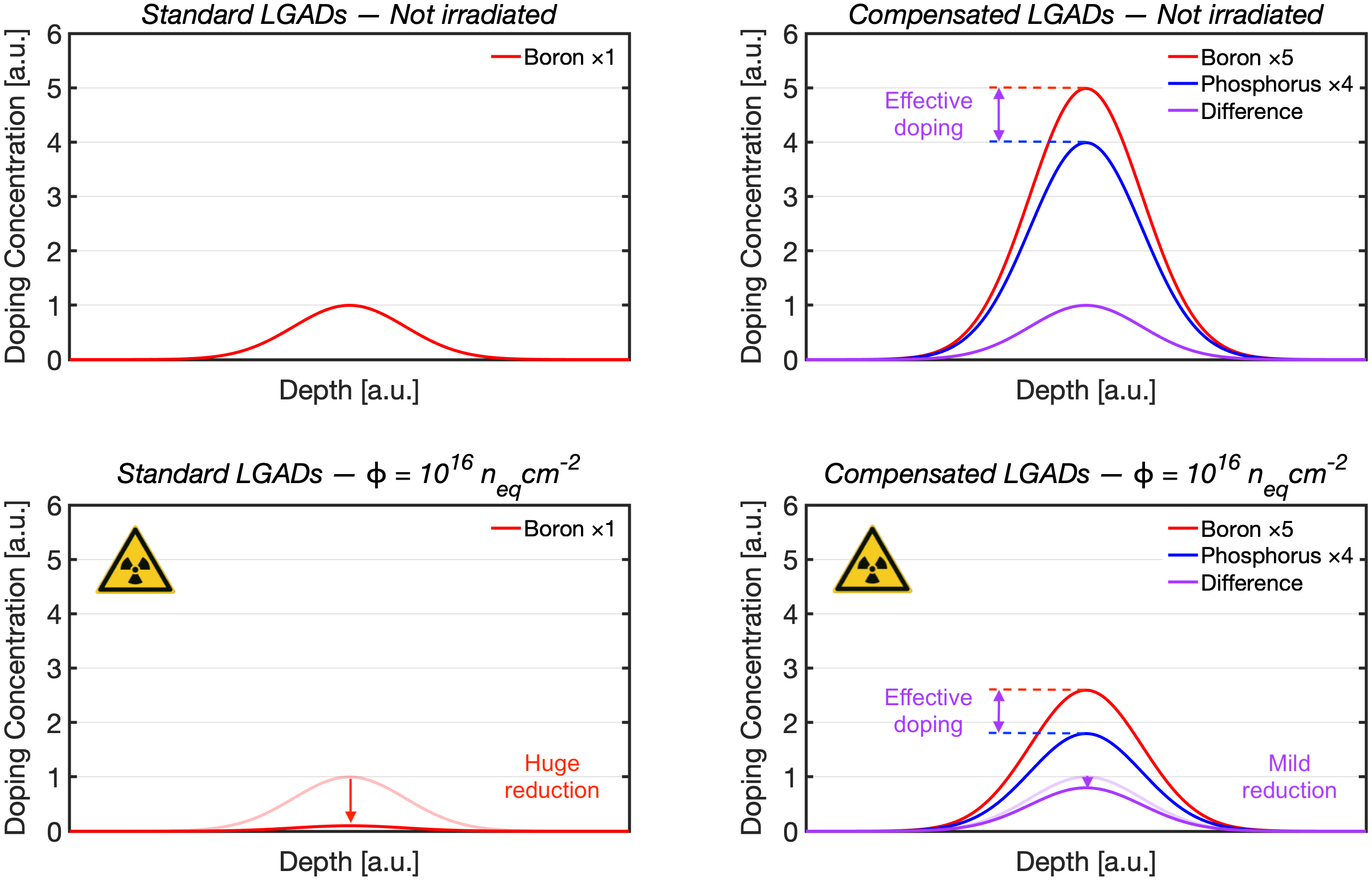}
        \caption{Gain implant evolution with irradiation in standard (left) and Compensated (right) LGADs. The top line refers to non-irradiated devices, and the bottom one to samples irradiated at $10^{16}$ $n_{eq}/cm^2$.}
        \label{fig:001}
    \end{figure}

    Low-gain avalanche diodes (LGADs)~\cite{PELLEGRINI201412} can provide the necessary timing resolution~\cite{CARTIGLIA201783} through internal signal amplification. However, their radiation resistance is inadequate for future hadron colliders. The gain implant, crucial for signal multiplication, deactivates under irradiation (Fig.~\ref{fig:001} left) through the acceptor removal mechanism~\cite{Kramberger2015}, degrading LGAD timing resolution. Most radiation-tolerant LGADs to date can survive up to $2.5\cdot10^{15}$ $n_{eq}/cm^2$~\cite{WHITE2024169798}, while the expected fluence in the FCC-hh’s innermost region will exceed $10^{17}$ $n_{eq}/cm^2$.

    A new gain implant design~\cite{Sola:2022dyc}, achieved through the compensation of two dopants of opposite type (Fig.~\ref{fig:001} right), has recently been introduced to enhance the radiation resistance of LGADs to extreme fluences. Both implants will undergo doping removal with irradiation; if engineered correctly, their difference will remain fairly constant, potentially extending the lifespan of Compensated LGADs beyond $10^{17}$ $n_{eq}/cm^2$.

\section{First production of Compensated LGADs}
\label{sec:FirstProduction}

    The first Compensated LGADs were manufactured by Fondazione Bruno Kessler (FBK) in late 2022 on 30 µm thick high-resistivity active substrates~\cite{Sola:2024whl}. Different combinations of boron (p$^\text{+}$) and phosphorus (n$^\text{+}$) doses have been investigated for the gain implant, as reported in the split table~\ref{tab:001}. Furthermore, carbon has been co-implanted in one wafer to investigate its effect in slowing acceptor removal when also phosphorus atoms are present in the same volume of the boron.

    The samples have been extensively characterised before and after irradiation\footnote{Sensor irradiation was done at the JSI TRIGA Mark II neutron reactor in the fluence range $[4\cdot10^{14},5\cdot10^{15}] \ n_{eq}/cm^2$.} with electrical and transient measurements~\cite{Sola:2024whl}. From I-V characteristics\footnote{I-V measurements before irradiation have been made at +20 ºC, whilst characterisations after irradiation have been performed at -20 ºC.}, it has been observed that the increase of the current with bias, due to the internal multiplication, seems to be maintained up to the highest fluences, meaning a higher radiation resistance compared to standard LGADs. Moreover, a timing resolution around 40 ps has been measured with samples irradiated at $2.5\cdot10^{15}$ $n_{eq}/cm^2$, indicating that Compensated LGADs can reach timing performances similar to the standard LGADs.

    \begin{table}
        \centering
        \begin{tabular}{cccc}
            \toprule
                Wafer No & p$^\text{+}$ dose & n$^\text{+}$ dose & C dose \\
            \midrule
                5 & 1 & {} & 1 \\
            \cmidrule{2-4}
                6 & 2a & 1 & {}\\
                7 & 2b & 1 & {}\\
                8 & 2b & 1 & {}\\
                9 & 2c & 1 & {}\\
            \cmidrule{2-4}
                10 & 3a & 2 & {}\\
                11 & 3b & 2 & {}\\
                12 & 3b & 2 & {}\\
                13 & 3b & 2 & 1\\
                14 & 3c & 2 & {}\\
            \cmidrule{2-4}
                15 & 5a & 4 & {}\\
            \bottomrule
        \end{tabular}
        \caption{Split table of the first Compensated LGADs manufactured by FBK in late 2022. For the dose coding, a $<$ b $<$ c and 2c $<$ 3a. Furthermore, W5 is a standard LGAD reported for reference.}
        \label{tab:001}
    \end{table}

\section{TCAD investigation}
\label{sec:TCAD}

    Comparing experimental measurements with TCAD simulations can provide additional insights for designing the next batch of optimised Compensated LGADs. In detail, the donor removal rate at high initial donor concentrations ($>$ $10^{16}$ at/cm$^3$) used in LGADs can be assessed, the interplay between acceptor and donor removal can be studied, and lastly, the impact of donor doping on the beneficial effect of carbon in slowing down acceptor removal can be evaluated.

    To this end, the state-of-the-art Synopsys\textsuperscript \textregistered\ Sentaurus TCAD suite has been used, and the radiation damage has been taken into account by using the last release of the Perugia radiation damage model~\cite{Morozzi:2024KG}. In the latter, acceptor and donor removal are parametrised accordingly to the following equation~\cite{MOLL2000282}:
    \begin{equation}
        N_{A,D}(\phi) = N_{A,D}(0) \cdot e^{-c_{A,D} \cdot \phi}
    \end{equation}
    where $c_{A,D}$ is the acceptor (donor) removal coefficient depending on the initial acceptor (donor) concentration $N_{A,D}(0)$, and $\phi$ is the irradiation fluence in $n_{eq}/cm^2$. The $c_A$ values are known across a wide range of $N_A(0)$ based on studies of standard LGADs~\cite{FERRERO201916}, while the literature on $c_D$ values is limited to $N_D(0)$ up to $10^{14}$ at/cm$^3$~\cite{WUNSTORF1996228}.

    \subsection{C-V characteristics}
    \label{subsec:CV}

        As C-V characteristics are well known to carry information related to doping profiles, they have been selected to begin comparing measurements and simulations. Specifically, it has started with the C-Vs of wafer 12 (W12), following the methodology reported in~\cite{FONDACCI2024169811} and briefly summarised below:
        \begin{enumerate}
            \item Calibrate the substrate thickness and doping concentration in the TCAD environment using the C-V measurements of p-i-n diodes;
            \item Incorporate Gaussian fits from compensated gain implant SIMS into the simulated device to create a Compensated LGAD, then verify the agreement between C-V measurements and simulations before irradiation.
            \item Select the right $c_A$ for the given $N_{A_{peak}}(0)$\footnote{The C-V characteristics are sensitive to changes in the peak concentration of the active gain implant~\cite{Croci:2022ygd}. Therefore the $c_A$ ($c_D$) of the peak is used to scale down the entire boron (phosphorus) profile.} from the acceptor removal parametrisation~\cite{FERRERO201916} and vary $c_D$ until C-V measurements and simulations agree after irradiation.
        \end{enumerate}

        The removal coefficients that enable the simulations to align with the measurements are $c_D = 6.50 \cdot 10^{-16}\ cm^2$ and $c_A = 2.50 \cdot 10^{-16}\ cm^2$~\cite{FONDACCI2024169811}, and $c_D \sim 2 \cdot c_A$ assuming the same dependence on the initial doping density and considering the $3-2$ initial concentration of implanted boron and phosphorus.

\begin{figure}[h]
            \centering
            \includegraphics[width=0.9\linewidth]{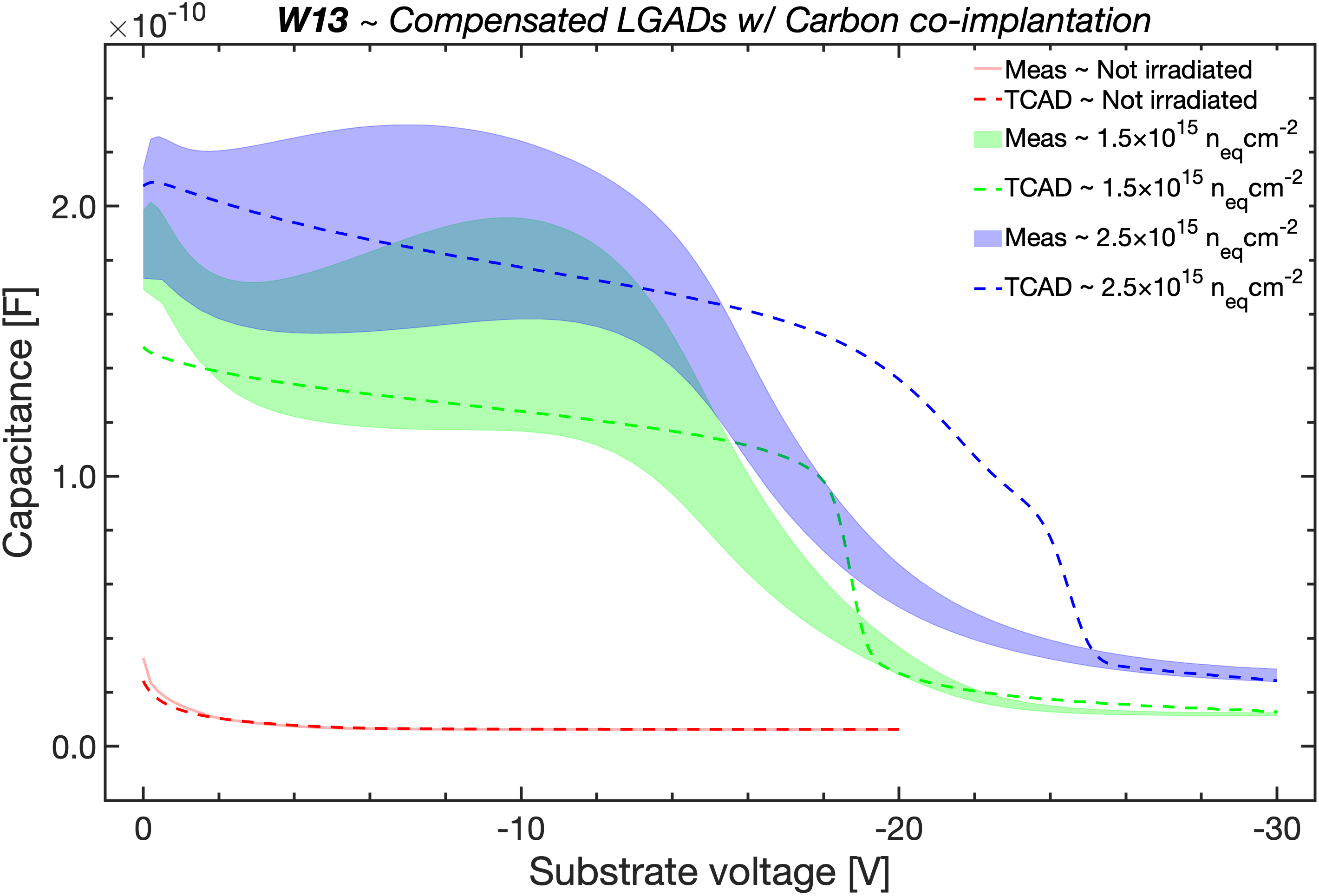}
            \caption{Pre- and post-irradiation W13 C-V measurements and simulations. Each band is the plane area covered by the family of curves obtained by measuring different samples of the same wafer under the same conditions.}
            \label{fig:003}
        \end{figure}

        Moving to the W13, which is similar to W12 but features carbon co-implantation, a good agreement between measurements and simulations (Fig.~\ref{fig:003}) was obtained with the previously extracted $c_D$ and with $c_A = 8.26 \cdot 10^{-17}\ cm^2$. The latter being one-third of the previous $c_A$ confirms the experimental observation~\cite{Sola:2024whl} that carbon slows acceptor removal in the same way as in standard LGADs~\cite{FERRERO201916}, even in the presence of phosphorus in the gain implant region.

    \subsection{Sheet resistance}
    \label{subsec:vdP}

        A new method for investigating doping removal by observing changes in sheet resistance with irradiation has been explored. Consequently, measurements and simulations of van der Pauw test structures~\cite{vdP} have been compared. These four-terminal structures, commonly utilised by foundries to monitor process quality, allow for the measurement of the sheet resistance of a given dopant layer by applying a known current between two terminals and probing the voltage drop across the other two.

\begin{figure}[h]
            \centering
            \includegraphics[width=0.9\linewidth]{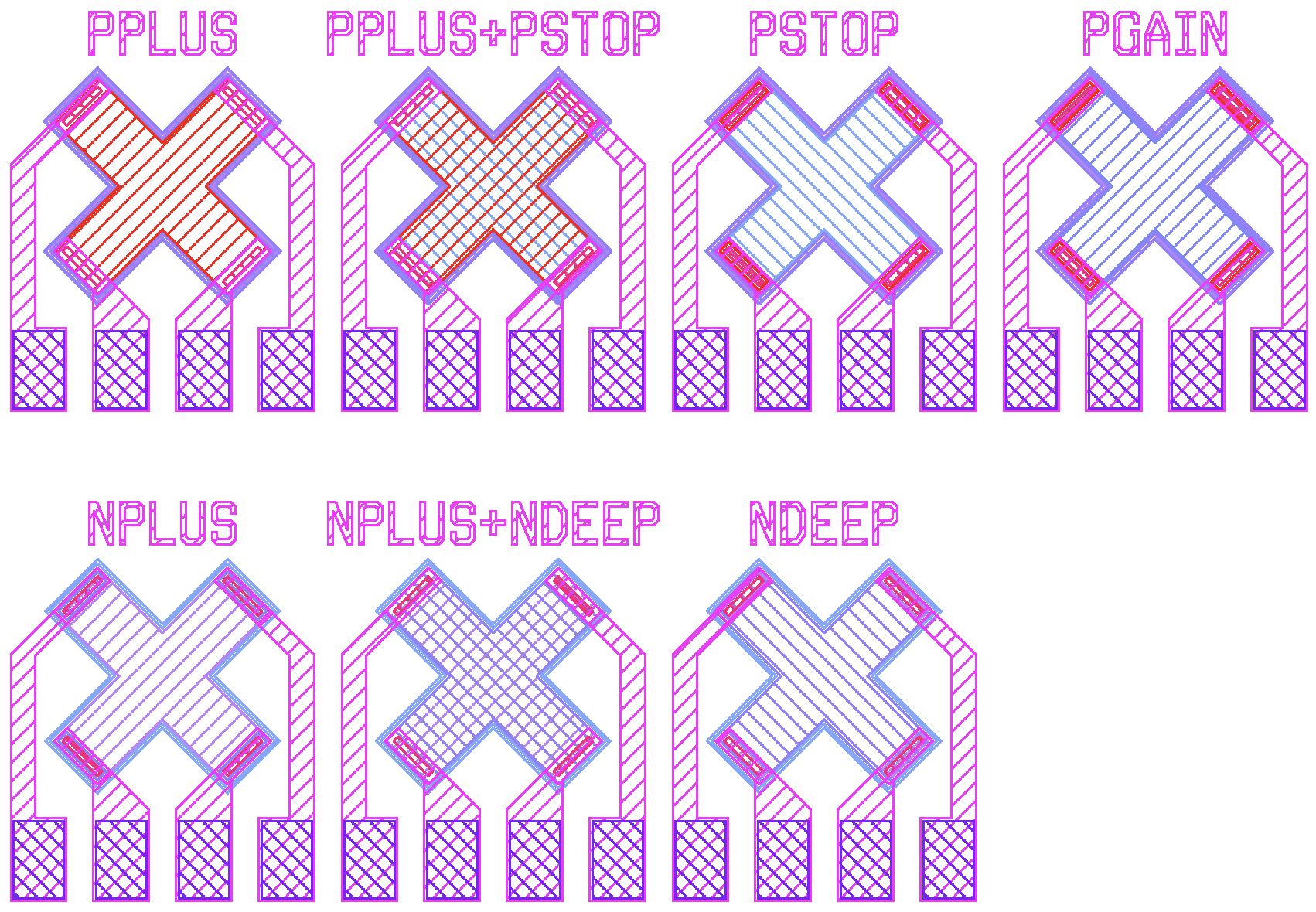}
            \caption{Layout of the van der Pauw test structures included in the first batch of Compensated LGADs released by FBK in late 2022.}
            \label{fig:004}
        \end{figure}

        A van der Pauw test structure can be replicated in a 3D simulation domain, and the sheet resistance can be calculated by following the measurement procedure. Thus, the doping variation with irradiation can be determined by tuning the doping profile in the TCAD environment until it reproduces the experimental variation in sheet resistance with fluence. For completeness, a van der Pauw test structure for each doping implant used in Compensated LGADs was included in the batch (Fig.~\ref{fig:004}).

\begin{figure}[h]
            \centering
            \includegraphics[width=0.9\linewidth]{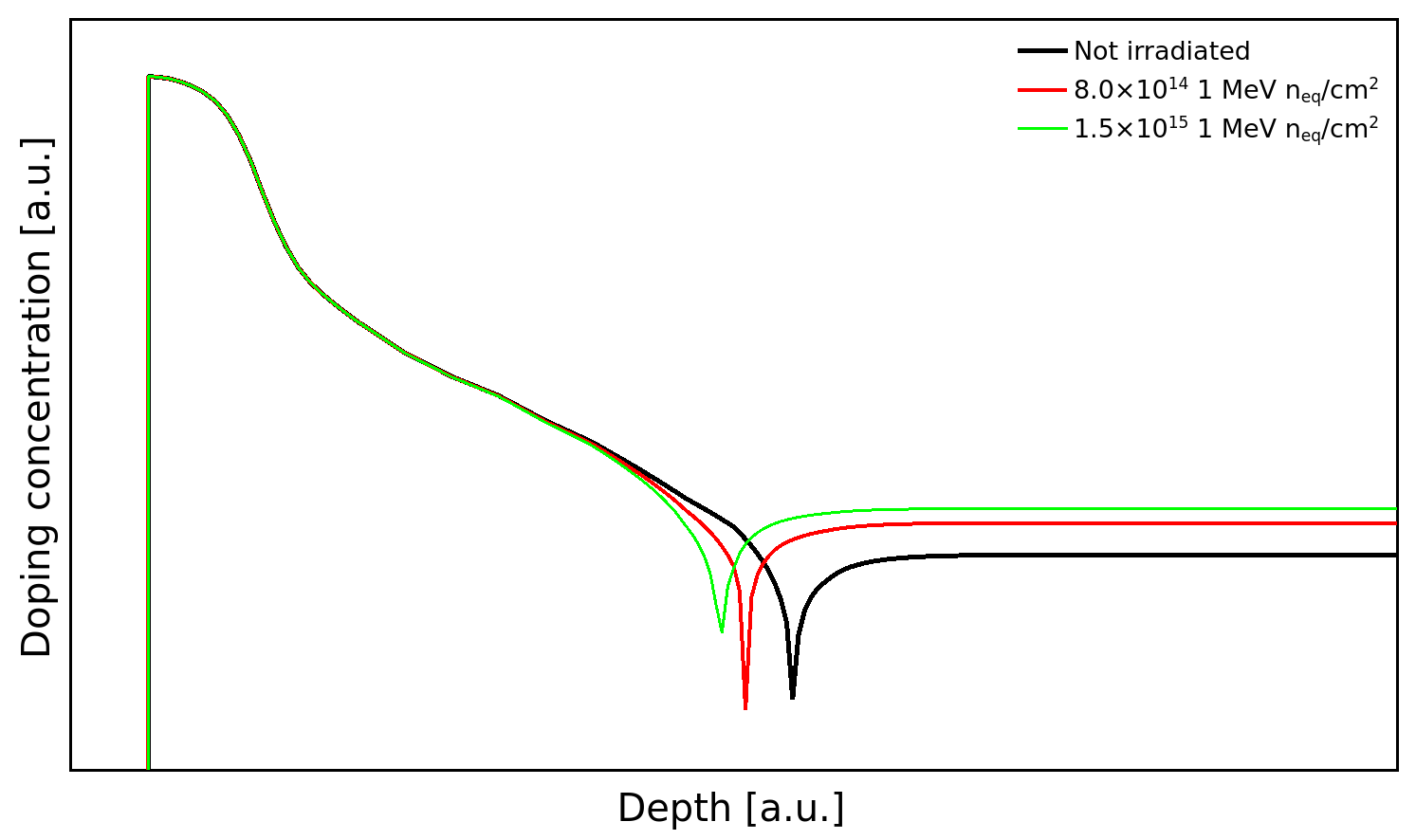}
            \caption{NPLUS doping profile evolution with irradiation. The substrate doping is increased accordingly to the acceptor creation parametrisation.}
            \label{fig:005}
        \end{figure}

\begin{figure}[h]
            \centering
            \includegraphics[width=0.9\linewidth]{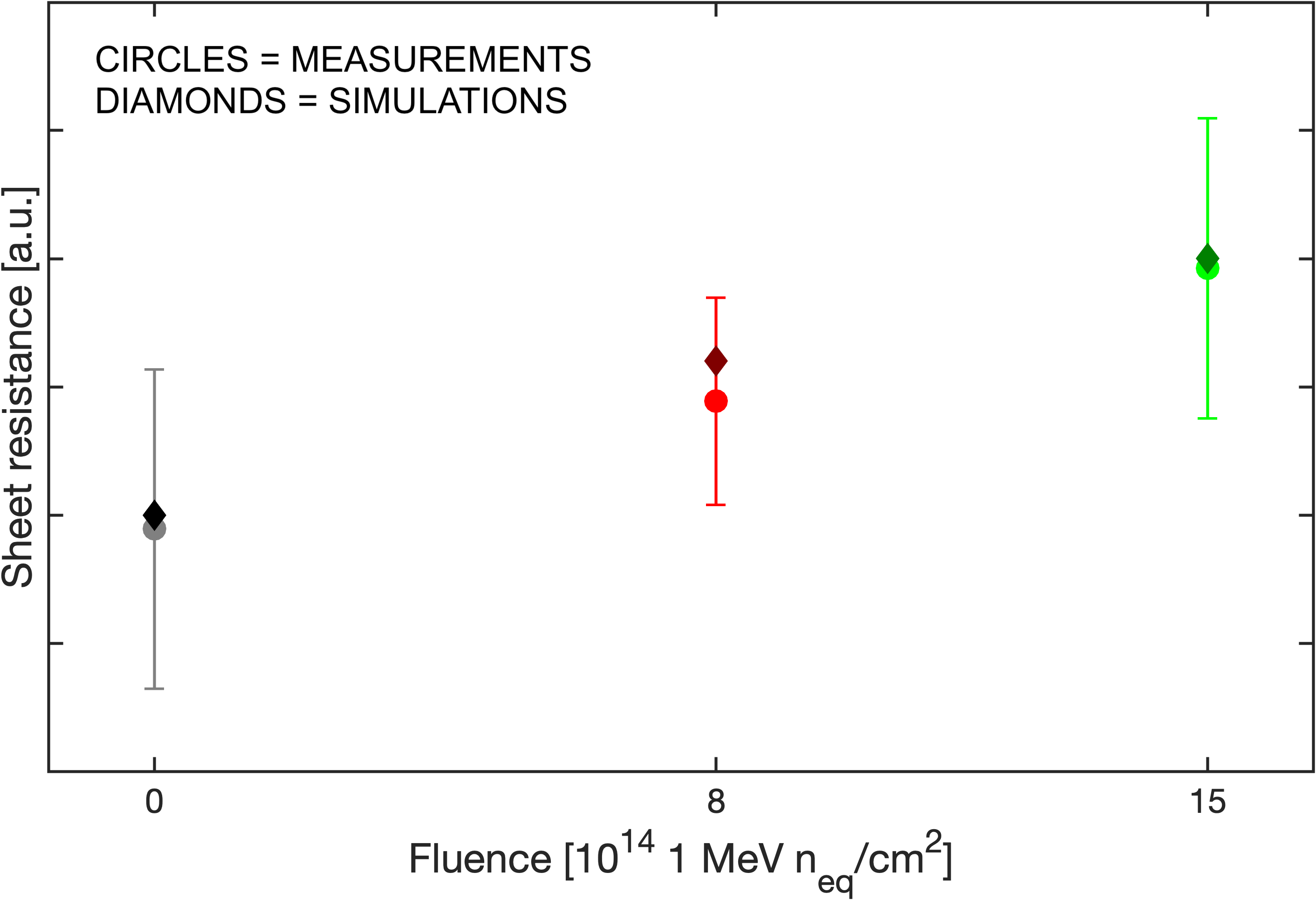}
            \caption{Pre- and post-irradiation sheet resistance measurements and simulations for the NPLUS implant. The error bars represent a small percentage of the average value, with their expansion caused by the scale being highly magnified.}
            \label{fig:006}
        \end{figure}

        Beginning with the NPLUS implant\footnote{Doping implant used by FBK to make the collection electrode in LGADs.}, Fig.~\ref{fig:005} illustrates the profile evolution which enables the reproduction of the experimental variation in sheet resistance with irradiation (Fig.~\ref{fig:006}). The black curve represents the non-irradiated scenario and depicts the NPLUS process simulation, calibrated on a SIMS, implanted into the p-type high-resistivity substrate. The black circle and diamond overlap in Fig.~\ref{fig:006} verifies that all the implanted NPLUS atoms are electrically active.

        The two after-irradiation profiles in Fig.~\ref{fig:005} result from increasing the substrate doping concentration based on the acceptor creation parametrisation~\cite{FERRERO201916}, thereby reducing the NPLUS tail. This reproduces the sheet resistance variation with irradiation (Fig.~\ref{fig:006}), demonstrating that the NPLUS implant is not affected by donor removal and confirming the quality of the acceptor creation parametrisation embedded in the Perugia radiation damage model.

        Considering a p-type implant on a p-type substrate, such as the PGAIN of W5, the substrate contributes to the measured sheet resistance with a parasitic effect, as the studied layer and the substrate are not separated by a depletion region. However, the simulation can reproduce this parasitic contribution; therefore, valuable information can be obtained by comparing measurements and simulations, even in this case.

        \begin{figure}[h]
            \centering
            \includegraphics[width=0.9\linewidth]{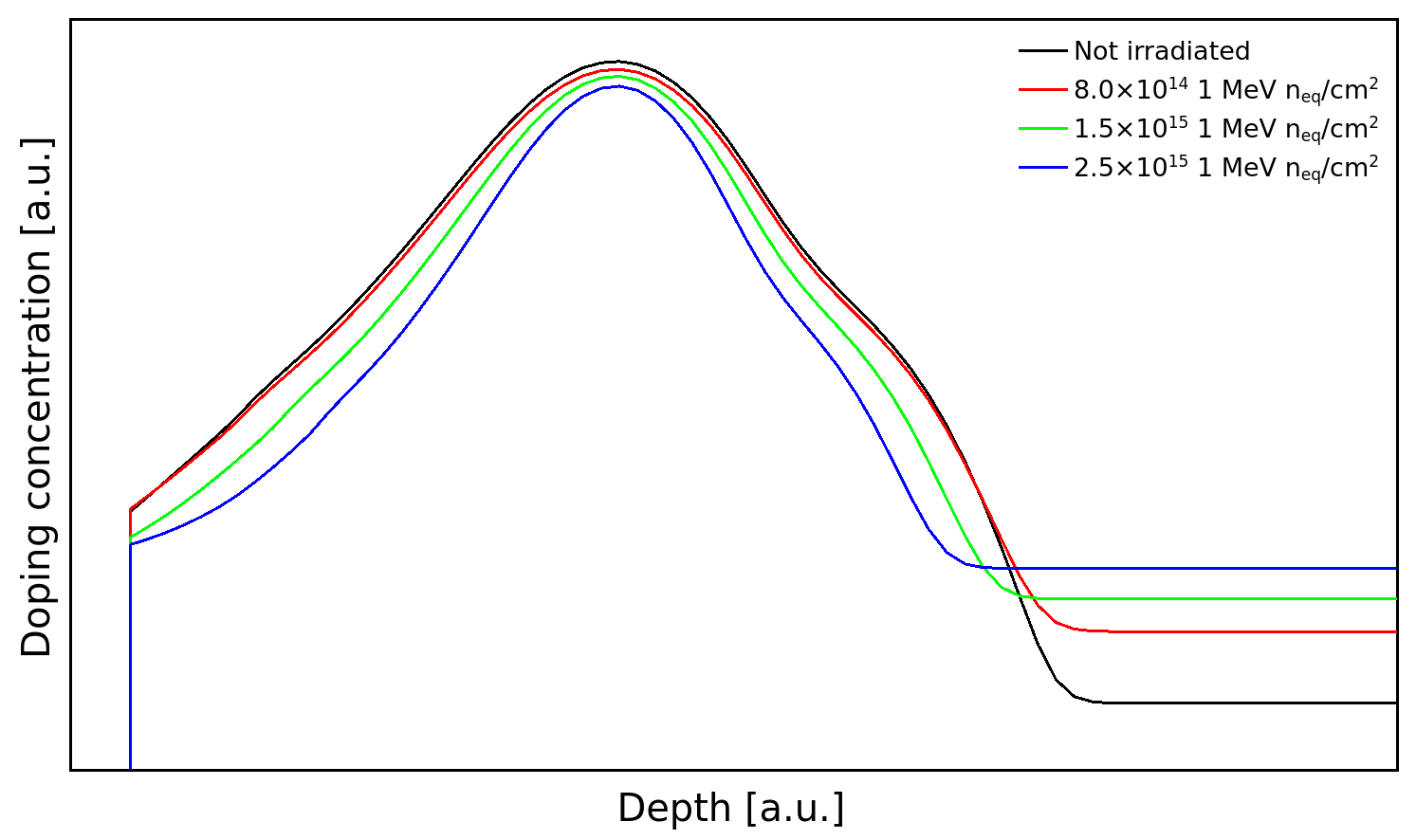}
            \caption{W5 PGAIN doping profile evolution with irradiation. The substrate doping increases following the acceptor creation parametrisation, while the gain implant doping decreases accordingly to the acceptor removal parametrisation.}
            \label{fig:007}
        \end{figure}

\begin{figure}[h]
            \centering
            \includegraphics[width=0.9\linewidth]{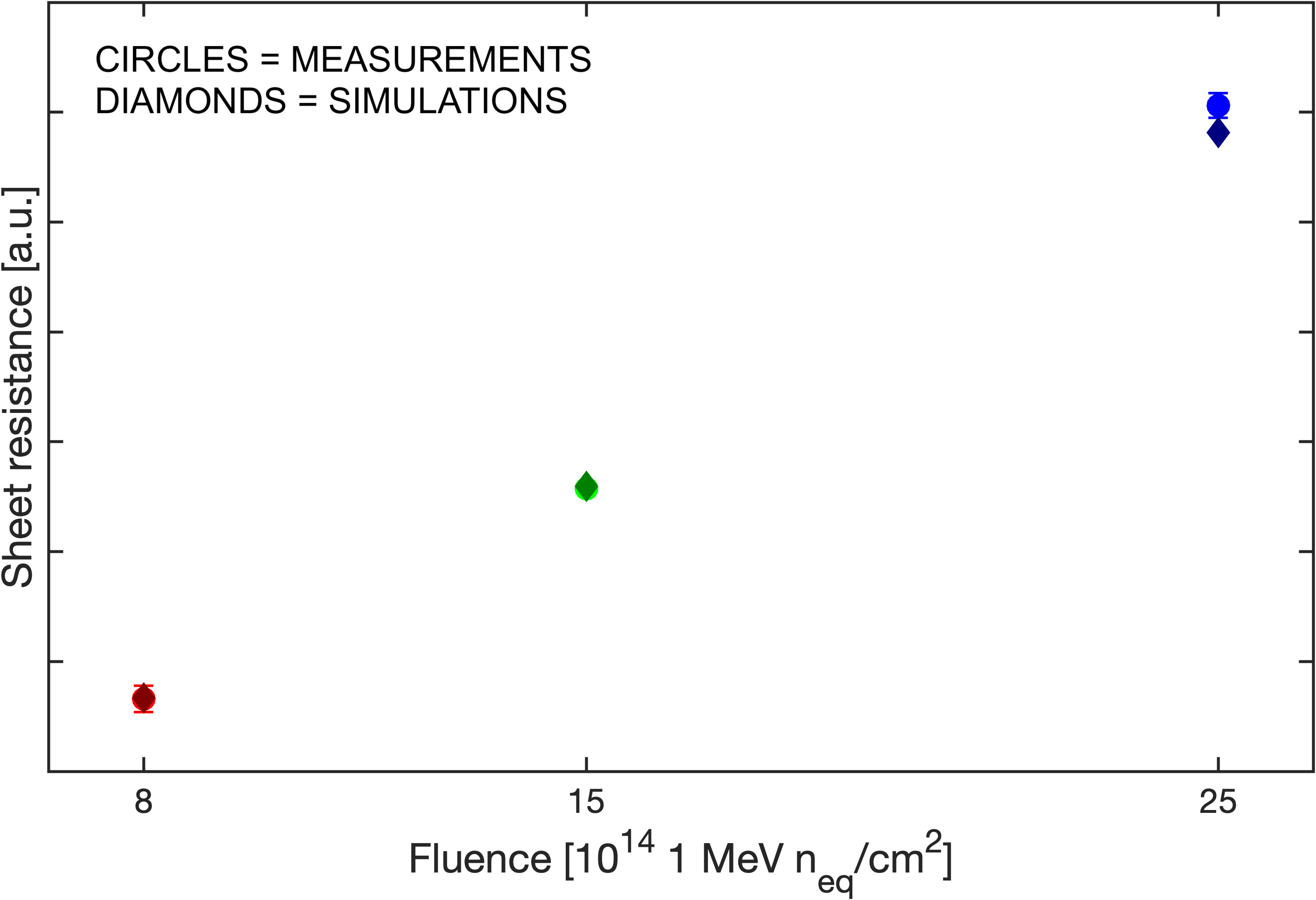}
            \caption{Pre- and post-irradiation sheet resistance measurements and simulations for the W5 PGAIN implant.}
            \label{fig:008}
        \end{figure}

        Fig.~\ref{fig:007} shows the evolution of W5’s PGAIN profile with irradiation, which allows for the reproduction of the changes in the experimental sheet resistance (Fig.~\ref{fig:008}). The black curve illustrates a process simulation calibrated using SIMS, while the post-irradiation curves have been derived by applying the acceptor removal (gain implant) and creation (substrate) parametrisations to the former. In particular, to reproduce the experimental data, the appropriate $c_A$ for each gain implant point had to be used in the acceptor removal parametrisation, and it was not possible to use the $c_A$ of the peak for all of them, resulting in more scaled tails. This indicates that the measurement of sheet resistance is more sensitive to changes in the integral of the active gain implant concentration, which also considers the reduction of profile tails, rather than variations in peak concentration only, as is the case for C-V characteristics.

\section{Conclusion}
\label{sec:Conclusion}

    The characterisation of the first Compensated LGAD prototypes built by FBK in 2022, which utilise the compensation of two opposite-type dopants for the gain implant, showcased their potential for extending timing measurements to extreme fluences ($>10^{17}$ $n_{eq}/cm^2$) anticipated in future hadronic colliders like FCC-hh.

    Moreover, by comparing experimental data with TCAD simulations before and after irradiation, doping removal due to irradiation has been investigated, offering insights for future Compensated LGAD batches. The Perugia radiation damage model was applied within the TCAD environment to accurately assess irradiation effects.

    Precisely, the measurement-simulation comparison of C-V characteristics facilitated the study of donor removal at high initial donor concentrations ($>10^{16}$ at/cm$^3$) used in Compensated LGADs, estimating a donor removal rate about twice that of acceptor removal, assuming the same functional form for both mechanisms. It was also confirmed that donor doping in the compensated gain implant does not affect the benefits of carbon co-implantation in slowing acceptor removal.

    Lastly, for the first time, measurements and simulations of van der Pauw test structures before and after irradiation have been compared to validate the change in sheet resistance as a method for assessing doping removal. This approach can independently estimate the variation of each doping implant used in Compensated LGADs, utilising a dedicated van der Pauw test structure for each of them. For instance, analysis of the NPLUS implant test structure demonstrated insensitivity to donor removal due to its high initial phosphorus concentration. Given its potential, this methodology will be applied to the new n-type LGAD batch currently in production at FBK.

\section*{Declaration of competing interest}
\label{sec:Declaration}

    The authors declare that they have no known competing financial interests or personal relationships that could have appeared to influence the work reported in this paper.

\section*{Acknowledgments}
\label{sec:Acknowledgments}

    This project has received funding from the European Union’s Horizon 2020 Research and Innovation Programme under GAs Nos 101004761 (AIDAinnova) and 101057511 (EURO-LABS), the PRIN MUR project 2022RK39RF ‘ComonSens’, and the European Union (ERC, CompleX, 101124288). Views and opinions expressed are however those of the authors only and do not necessarily reflect those of the European Union or the European Research Council. Neither the European Union nor the granting authority can be held responsible for them.

\bibliography{mybibfile}

\end{document}